\documentclass[letter]{aa} 
\usepackage{color}
\usepackage{ctable}
\usepackage{natbib,twoopt}
\usepackage{multirow}
\usepackage{amsmath,amssymb}
\usepackage{threeparttable}
\usepackage[breaklinks=true]{hyperref} %% to avoid \citeads line fills
\bibpunct{(}{)}{;}{a}{}{,} %% natbib format for A&A and ApJ
\makeatletter
\newcommandtwoopt{\citeads}[3][][]{\href{http://adsabs.harvard.edu/abs/#3}%
{\def\hyper@linkstart##1##2{}%
\let\hyper@linkend\@empty\citealp[#1][#2]{#3}}}
\newcommandtwoopt{\citepads}[3][][]{\href{http://adsabs.harvard.edu/abs/#3}%
{\def\hyper@linkstart##1##2{}%
\let\hyper@linkend\@empty\citep[#1][#2]{#3}}}
\newcommandtwoopt{\citetads}[3][][]{\href{http://adsabs.harvard.edu/abs/#3}%
{\def\hyper@linkstart##1##2{}%
\let\hyper@linkend\@empty\citet[#1][#2]{#3}}}
\newcommandtwoopt{\citeyearads}[3][][]%
{\href{http://adsabs.harvard.edu/abs/#3}
{\def\hyper@linkstart##1##2{}%
\let\hyper@linkend\@empty\citeyear[#1][#2]{#3}}}
\makeatother
\usepackage{hyperref}

\usepackage{txfonts}
\def\Teff{\ensuremath{T_{\mathrm{eff}}}}

\def\vsini{\ensuremath{{\upsilon}\sin i}}

\def\kms{$\mathrm{km\,s}^{-1}$}

\def\logl{\ensuremath{\log (L/{L_\odot})}}

\def\M{\ensuremath{M_{\odot}}}

\def\Pt{\ensuremath{\rm P_{\mathrm{turb}}\,}}
\def\Ptmax{\ensuremath{\rm P_{\mathrm{turb}}^{\mathrm{max}}\,}}

\def\Msun{\ensuremath{\rm \,M_\odot\,}}

\def\logLl{\rm log(L/L_{\odot})}
\def\vmac{$\upsilon_{\mathrm{mac}}$}

\begin{document} 
\title{Relating turbulent pressure and macroturbulence across the HR diagram with a possible link to $\gamma$ Dor stars}
\author{L. Grassitelli\inst{1},
        L. Fossati\inst{2,1},
        N. Langer\inst{1},
        A. Miglio\inst{3},
        A.~G. Istrate\inst{1},
        D. Sanyal\inst{1}
}
\institute{
        Argelander-Institut f\"ur Astronomie, Universit\"at Bonn, auf dem                         H\"ugel 71, 53121 Bonn, Germany\\
        \email{luca@astro.uni-bonn.de}
        \and
        Space Research Institute, Austrian Academy of Sciences, Schmiedlstrasse                 6, A-8042 Graz, Austria
        \and
        School of Physics \& Astronomy, University of Birmingham, Edgbaston,                         Birmingham, B15 2TT, UK
}
\date{} 
\abstract{A significant fraction of the envelope of low- and intermediate-mass stars is unstable to convection, leading to sub-surface turbulent motion. Here, we consider and include the effects of turbulence pressure in our stellar evolution calculations. In search of an observational signature, we compare the fractional contribution of turbulent pressure to the observed macroturbulent velocities in stars at different evolutionary stages. We find a strong correlation between the two quantities, similar to what was previously found for massive OB stars. We therefore argue that   turbulent pressure fluctuations    of finite amplitude may excite high-order, high-angular degree stellar oscillations, which manifest themselves at the surface an additional broadening of the spectral lines, i.e., macroturbulence, across most of the HR diagram. When considering the locations in the HR diagram where we expect high-order oscillations to be excited by stochastic turbulent pressure fluctuations, we find a close match with the observational $\gamma$ Doradus instability strip, which indeed contains high-order, non-radial pulsators. We suggest that turbulent pressure fluctuations on a percentual level may contribute to the $\gamma$ Dor phenomenon, calling for more detailed theoretical modeling in this direction.   
} 
\titlerunning{Turbulent pressure and macroturbulence broadening}
\authorrunning{L. Grassitelli et al.}
\maketitle
\section{Introduction}
%\label{sec:introduction}
%
Stars transport energy mostly via two physical mechanisms: radiative diffusion and convection. Convection consists of upward and downward turbulent motion of material and consequently leads, strictly speaking, to local deviation from hydrostatic equilibrium. \citet{2015Grassitelli} showed that the strength of turbulent pressure within the sub-surface convective zones (SSCZ) of hot massive stars may be related to the observationally derived macroturbulent velocities at the surface \citep{2014SimonDiaz}, as a consequence of the local lack of hydrostatic equilibrium at the percent level due to stochastic turbulent pressure fluctuations. 
However,  massive stars are not the only stars that show vigorous SSCZ. 

Turbulent convection is known to play a decisive role in the
stochastic forcing, self-excitation and damping of pulsation modes
across the HR diagram (see, e.g., \citealt{2000Houdek}, \citealt{2004Dupret}, \citealt{2007Xiong}, \citealt{2013Belkacem}. Moreover, a recent study by \citet{2014Antoci} showed that turbulent
pressure in the hydrogen ionization zone is likely to explain the
high-frequency, coherent pulsation modes observed in the $\delta$ Scuti
pulsator HD 187547. Hydrogen recombines at a temperature on the order of 10\,000\,K, leading to high values of the Rosseland opacity and to a high Eddington factor in the corresponding layers \citep{2015Sanyal}. This opacity bump could also play a role in the pulsational instability of the  $\gamma$ Doradus ($\gamma$ Dor) class of stars \citep{2002Handler}. 

Stars of the $\gamma$ Dor class are intermediate mass stars showing periodic variability on a timescale of days. These A- and F-type stars are pulsating with high-order, non-radial g-modes \citep{2000Guzik,2008Miglio}. Theoretical stability analyses adopting time-dependent convection predict an instability strip (IS) to occur close to where $\gamma$ Dor are observed, involving a flux-blocking excitation mechanism \citep{2000Guzik,2005Dupret}. However, \citet{2004Dupret} showed that the agreement between observations and theory depends very sensitively on the exact treatment of convection.

We investigate here the effects of turbulent pressure in intermediate mass stars, focusing on its observational fingerprints, and  try to draw a global picture of these effects across the HR diagram.
\section{Method}
%\label{sec:method}
%
We have computed a set of stellar evolutionary models using the hydrodynamic Lagrangian one-dimensional Bonn evolutionary code \citep{2000Heger,2006Yoon,2011Brott}. For convection the models are computed using the Ledoux criterion \citep{1990Kippenhahn}, adopting the mixing length theory (MLT) with a mixing length parameter $\alpha=1.5$ \citep{1958Vitense,2011Brott}. The opacities are computed from the OPAL opacity tables \citep{1996Iglesias}. No stellar wind mass loss was included in the calculations and we assumed a metallicity of Z\,=\,0.02 \citep{2011Brott}.

We also used for comparison a modified version of BEC, which includes the turbulent pressure (\Pt) and turbulent energy density ($\rm e_{turb}$) in the stellar structure equations \citep{2015Grassitelli}.
These quantities can be expressed as \citep{2003Stothers}
\begin{equation}\label{pturbeq}
P_{turb}= \zeta \rho v_c^2 \quad,\quad e_{turb}= \frac{3}{2} \frac{P_{turb}}{\rho} \quad ,
\end{equation}
where $\rho$ is the local density, $\zeta$ is a parameter chosen to be $\zeta=1/3$ for isotropic turbulence \citep{2003Stothers}, and $v_c$ is the local convective velocity. 

In Appendix A, Fig.~\ref{fig:1.5} shows an HR diagram comparing two 1.5\,\Msun tracks, with and without the inclusion of turbulent pressure and energy. We find that the structural effects arising from the inclusion of $\rm e_{turb}$ and \Pt do not significantly affect the evolutionary tracks, in agreement with \citet{2015Grassitelli}. Therefore, we  study the contribution of the turbulent pressure to the equation of state \emph{a posteriori}, using Eq.~\ref{pturbeq}. 
%-------------------------------------------

%-------------------------------------------
%
\section{Results}

%-------------------------------------------
\begin{figure}
\resizebox{\hsize}{!}{\includegraphics{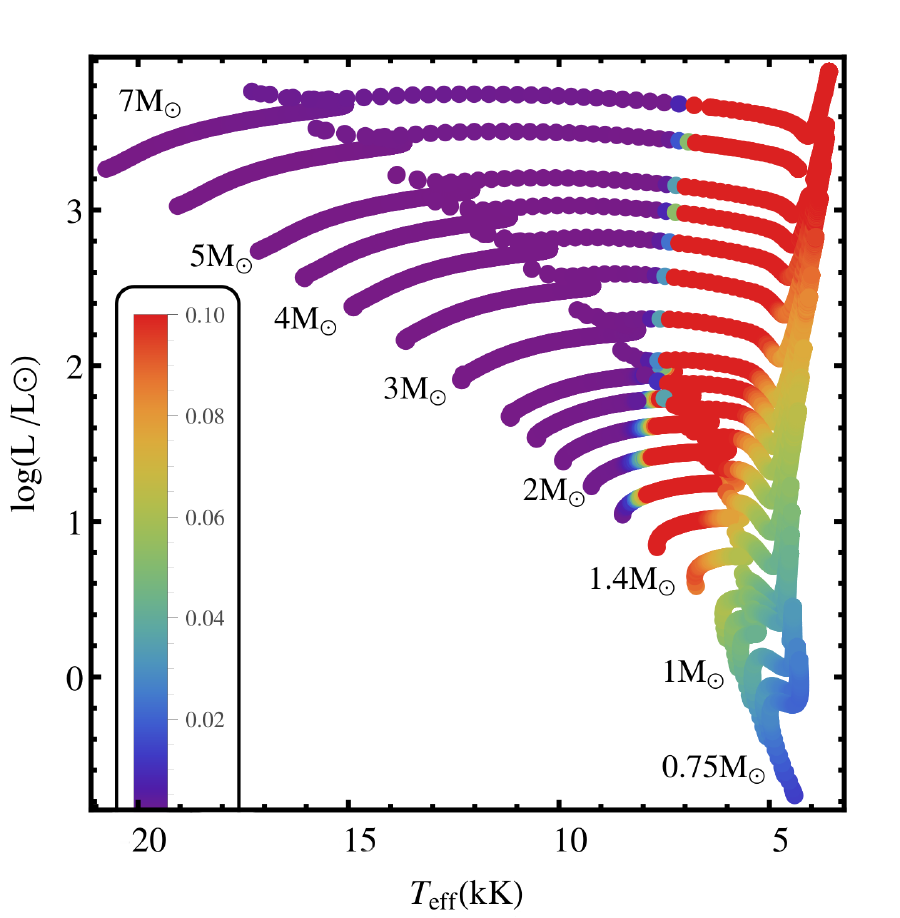}}
\caption{HR diagram showing the computed stellar evolution tracks. The colors of the dots indicate the maximum fraction of turbulent-to-total pressure occurring within the stellar envelopes (see framed color bar in the bottom left corner). Stellar models  in red show \Pt/P>10\%. The masses of the models are indicated next to the tracks. For the upper HR diagram, see Fig.1 in \citet{2015Grassitelli}.}
\label{fig:track}
\end{figure}
%-------------------------------------------

Figure~\ref{fig:track} shows the HR diagram with our newly computed evolutionary tracks for low- and intermediate-mass stars (0.75$-$7\Msun), color coded according to  the maximum ratio of turbulent-to-total pressure within each stellar model. 
At $\logLl\approx 1$ the models that show a significant \Pt\ fraction are confined within a rather narrow \Teff\ range of 6000--8000\,K, while at \Teff<8000\,K the \Pt\ contribution is small. Within this band, \Pt\ can account for up to $\approx$10--15\% of the total pressure. The hot boundary at \Teff\,$\approx$\,8000\,K coincides with a sharp increase in opacity very close to the surface owing to hydrogen recombination. The maximum fraction of turbulent pressure reaches a peak at \Teff$\sim$7500\,K, then followed  by a gradual decrease with decreasing \Teff, where the maximum fraction of turbulent pressure is reduced to $\approx$\,5\% at \Teff$\approx 4500\,$K. 

At high luminosity, e.g., $\logLl\approx 3$, the \Teff\ range within which the fraction of \Pt\ is larger than 10\% has moved to the interval 4000--7000\,K. This band forms a continuous region in the HR diagram with the higher luminosity red supergiants computed by \citet{2015Grassitelli}. Thus, it ranges from \logl\,$\approx$\,0.7, where it crosses the zero-age main-sequence (ZAMS) at M\,$\sim$\,1.5\,\M, up to the most luminous supergiants with \logl\,$\approx$\,5.5\, irrespective of the evolutionary stage. Within this band, \Ptmax\ goes from 10\% on the ZAMS to 33\% at high luminosities. 

\begin{figure}
\vspace{-0.5cm}
\resizebox{\hsize}{!}{\includegraphics{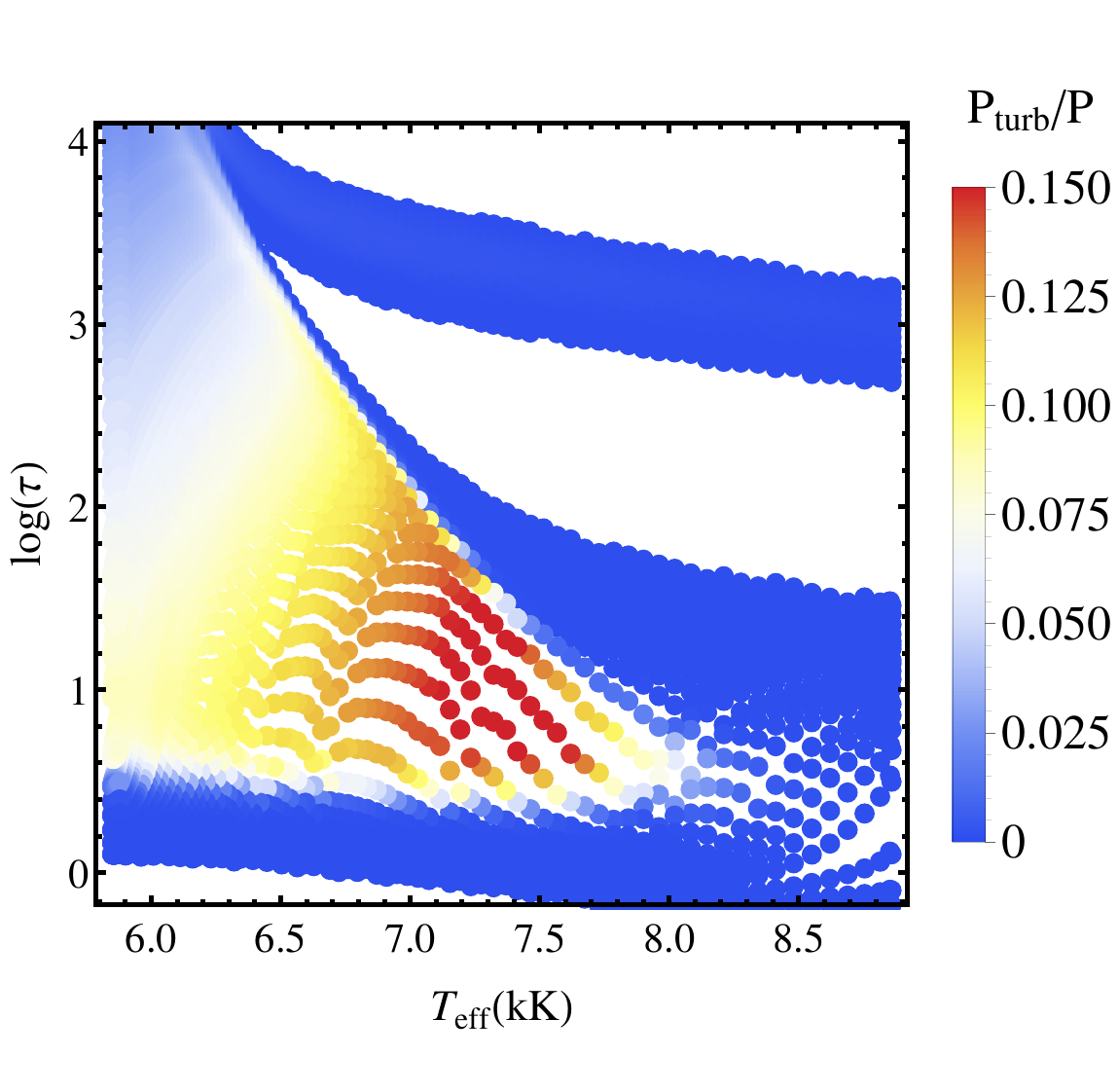}}\vspace{-0.5cm}
\caption{Ratio of turbulent-to-total pressure (color coded) as a function of the effective temperature and the optical depth throughout part of the evolution of a $1.9\Msun$ model. Extended white areas are radiatively stable. The HeCZ is visible in the top right section of the plot. For $\Teff>7700\,$K the HCZ is present from the surface up to $\rm log(\tau) \approx 2$. At lower \Teff\, the outer boundary of the HCZ is located at $\rm log(\tau) \approx 0$, while the inner boundary reaches $\rm log(\tau) > 3.5$ and merges with the HeCZ.
}
\label{fig:tefftau}
\end{figure}

The internal envelope structure can be seen in Fig.~\ref{fig:tefftau}, where the outer layers of the \,1.9\Msun model are shown for part of its evolution. 
Figure~\ref{fig:tefftau} shows that the hydrogen convective zone (HCZ) becomes more and more extended (in terms of optical depth) as the star cools, moving toward the giant branch, although its radial extent is only $\approx$\,5\% of the stellar radius for the coolest models. A steep increase of \Pt\ is present at \Teff \,$\approx8000$\,K as the inefficiency of convection causes the convective velocities to approach the sound speed
\citep[cf. ][]{2015Grassitelli}, followed by a smooth decrease as the superadiabaticity of the outer layers decreases (see Appendix~\ref{app:conv_times} for a discussion of convective timescales and mixing length parameter $\alpha$).
%-------------------------------------------
%-------------------------------------------

\section{Observational consequences of turbulence pressure}
\subsection{Macroturbulence broadening}\label{sec:vmac}
In the context of massive stars, \citet{2009Aerts} concluded that the joint effects of a large number of high-order, high-angular degree g-mode pulsations at the surface may mimic macroturbulent line broadening. \citet{2015Grassitelli} found that in hot massive stars macroturbulence broadening is strongly correlated to the maximum fraction of \Pt within the stellar models. They suggested that stochastic turbulent pressure fluctuations may excite high-order g-modes that then give  rise to the observed macroturbulence. We here consider the same possibility by looking at the same quantities at the opposite side of the HR diagram.

We collected observationally derived macroturbulent velocities (\vmac) for low- and intermediate-mass chemically normal stars previously derived from high resolution spectra by \citet{2008Carney}, \citet{fossati2011}, \citet{2014Doyle}, and \citet{Ryab2015}. We further looked for additional intermediate-mass chemically normal apparent slow rotators in order to increase the sample, particularly in the range of the evolved stars. Given the relatively small \vmac\ values (typically $\sim$10\,\kms), all stars for which it is possible to reliably measure \vmac\ have a low projected rotational velocity (\vsini) of at most 10--15\,\kms. We therefore derived macroturbulent velocities for HD\,167858 \citep{bruntt2008}, KIC\,8378079, KIC\,9751996, and KIC\,11754232 \citep{2015Reeth}, all reported to have a \vsini\ below 15\,\kms.

 For all stars, we analyzed the spectral lines as described in \citet{fossati2011}, and we adopted the atmospheric parameters listed in \citet{bruntt2008} for HD\,167858 and in \citet{2015Reeth} for KIC\,8378079, KIC\,9751996, and KIC\,11754232. The results, listed in Table~\ref{tab:vsini-vmac}, show that for all stars we obtained a non-negligible \vmac\ value. We also attempted to fit the spectral lines not accounting for macroturbulent line broadening, which resulted in significantly worse fits compared to the case with macroturbulent broadening included. 

\begin{figure}
\vspace{-0.45cm}
\resizebox{\hsize}{!}{\includegraphics{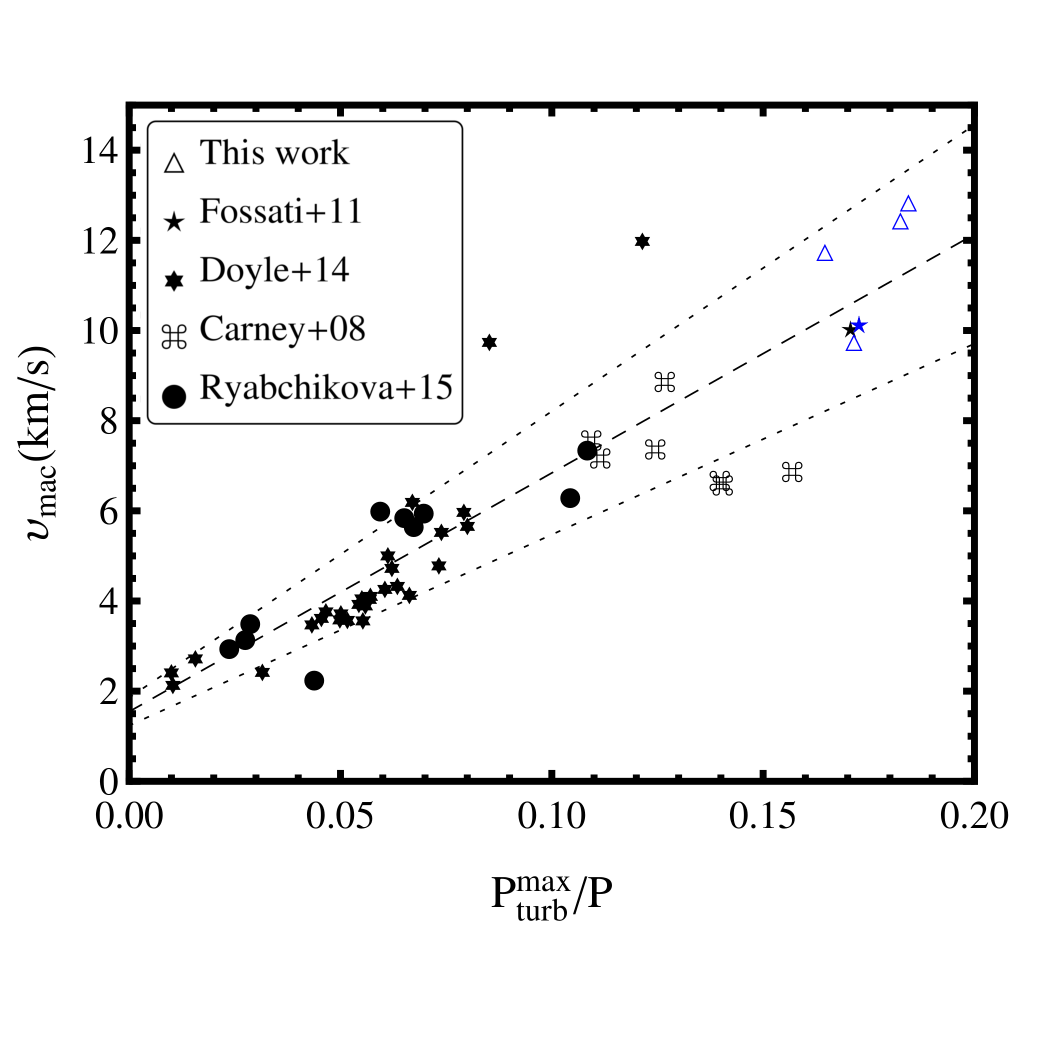}}
\vspace{-1.3cm}
\caption{Observed macroturbulent velocities as a function of the estimated fraction of turbulent pressure from our calculations (via a fitting of the tracks). The data are from \citet{2008Carney}, \citet{fossati2011}, \citet{2014Doyle}, \citet{Ryab2015}, and this work. The dotted lines represent typical errors (i.e., 20\% of \vmac, equivalent to 1 standard deviation) inferred to the linear fit (dashed line). The five stars color coded in blue are bona fide $\gamma$ Dor pulsators.}
\label{fig:vmac-Pturb}
\end{figure}

\begin{figure*}
\begin{sidecaption}
\resizebox{0.70\hsize}{!}{
\includegraphics{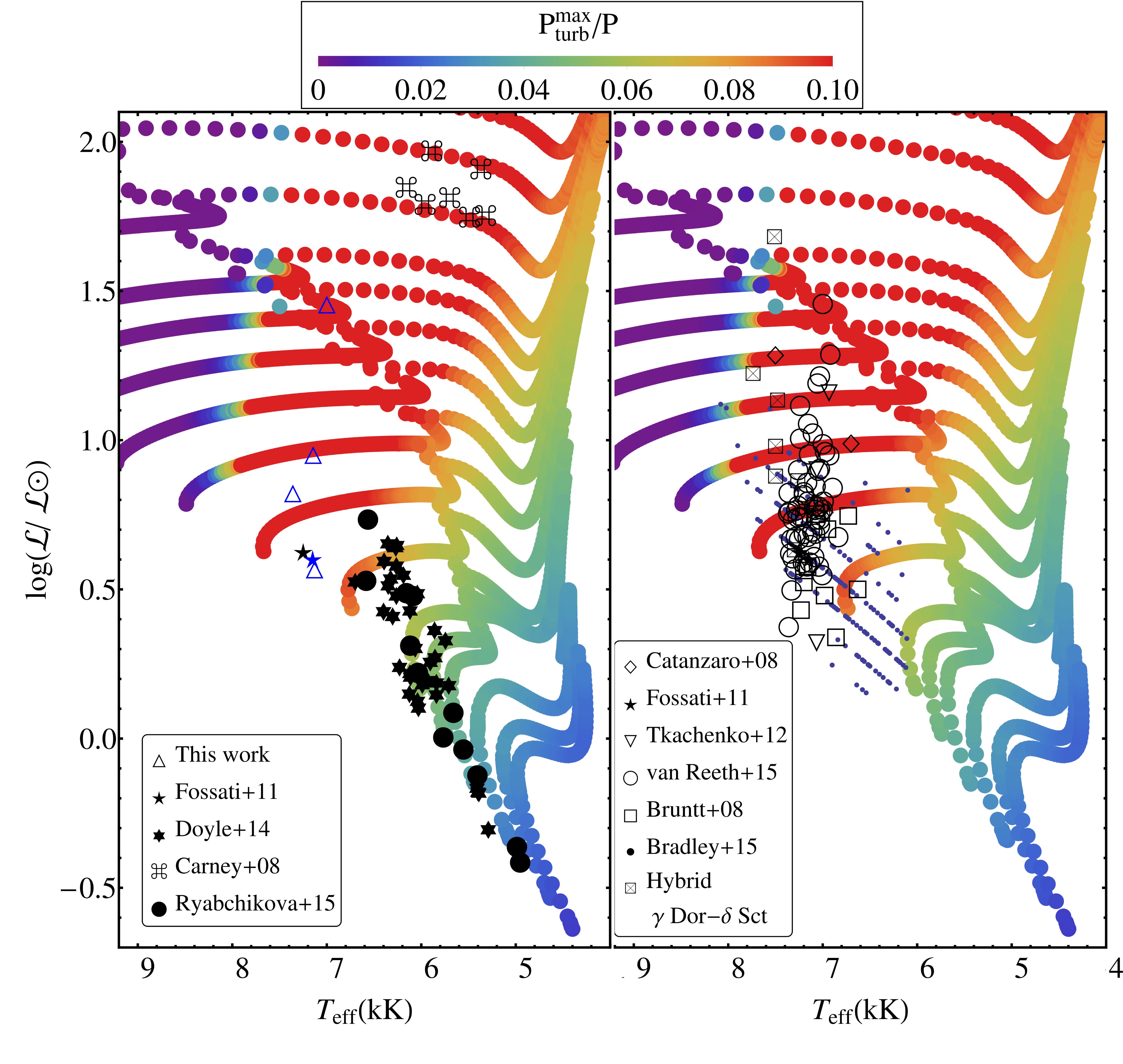}}\vspace{-0.6cm}
\caption{
Left: stellar evolutionary tracks in a sHR diagram color coded according to the maximum fraction of \Pt/P (see top bar). Overplotted are the location of 56 stars analyzed by \citet{2008Carney}, \citet{fossati2011}, \citet{2014Doyle}, and \citet{Ryab2015} for which macroturbulent velocities are available, in addition to those listed in Tab. \ref{tab:vsini-vmac}. The five stars color coded in blue are bona fide $\gamma$ Dor pulsators.\newline
Right: As in  left panel, with overplotted the location of $\approx300$ $\gamma$ Dor stars analyzed by \citet{bruntt2008}, \citet{2011Catanzaro}, \citet{fossati2011}, \citet{2012tkachenko}, \citet{2015Bradley}, and \citet{2015Reeth} and some hybrid $\gamma$ Dor--$\delta$ Sct stars from \citet{2011Catanzaro} and \citet{2012tkachenko}.
\vspace{+4cm}
}
\label{fig:sHR}
\end{sidecaption}
\end{figure*}

Figure~\ref{fig:vmac-Pturb} shows the macroturbulent velocities as a function of the maximum fraction of turbulent pressure obtained from the stellar tracks at the position in the HR diagram of each observed star. The left panel of Fig.~\ref{fig:sHR} shows the position of the stars considered here in the spectroscopic HR diagram \citep[sHR;][]{2014Langer} superposed on the computed stellar tracks. 

Similar to the case of the massive stars \citep[see Fig.~5 of][]{2015Grassitelli}, we obtain a strong correlation between the two quantities: macroturbulent velocities increase with increasing \Ptmax/P. The relation is highly significant, with a Spearman-rank correlation coefficient of  0.96.  In chemically normal slowly rotating MS late B- and A-type stars the spectral lines do not show the presence of macroturbulent broadening \citep[e.g.,][]{fossati2009}, and  the stellar models consistently predict extremely small \Ptmax in that region.  
Moreover, \citet{gray1984}, \citet{2005Valenti}, and \citet{Ryab2015} find a trend of increasing macroturbulent velocities for increasing effective temperatures in a sample of cool stars in the range 4500\,K<\Teff<6500\,K. This is  in a good agreement with our prediction as well, given that we expect an increase of \Pt with \Teff\ in that temperature range (see Sect.3 and Fig.\ref{fig:tefftau}), providing additional evidence of a relation between \vmac\ and \Pt.

%-------------------------------------------
%-------------------------------------------
%-------------------------------------------
%-------------------------------------------
%
\subsection{The $\gamma$ Dor instability strip}
Our results point toward the presence of a strong correlation between macroturbulent broadening and turbulent pressure across the whole HR diagram. In this picture, macroturbulent broadening may be the result of high-order modes excited by turbulent pressure fluctuations \citep{2015Grassitelli}. Interestingly, the region of high \Pt\ shown in Fig.~\ref{fig:track} broadly overlaps with the theoretical instability strips of $\gamma$ Dor stars, which are indeed high-order, g-mode pulsators \citep{2011Bouadib}.

To further explore the possible \Pt-$\gamma$ Dor connection,  in the right panel of Fig.~\ref{fig:sHR} we compare the position in the sHR diagram of observed $\gamma$ Dor stars and hybrid $\gamma$ Dor -- $\delta$ Sct stars with the maximum value of \Pt/P as derived from the stellar models. This comparison shows that the observational $\gamma$ Dor instability strip falls in the same area of the sHR diagram where we find the largest contribution of \Pt\ to the total pressure. In addition, the stars that present a non-negligible macroturbulent broadening lie predominantly along this same area of the sHR diagram. As a matter of fact, all the stars listed in Tab.~\ref{tab:vsini-vmac} and HD\,37594 are indeed bona fide $\gamma$ Dor pulsators \citep{diaz2014,2015Reeth}.

If turbulent pressure fluctuations act as an excitation mechanism for high-order pulsation modes, it may well be that turbulent pressure fluctuations of finite amplitude play a role in the rise of the $\gamma$ Dor phenomenon. Rigorous studies of the Reynolds stress tensor in connection with time-dependent convection \citep{2005Grigahcene,2014Antoci} and eventually finite amplitude perturbations \citep{2015Grassitelli} are required  to investigate the physical origin of the possible connection between high-order g-modes, \vmac, and sub-surface turbulent pressure fluctuations.

\section{Conclusions}

We investigate the effects of convective turbulent pressure in late-type low- and intermediate-mass stars, focusing on their possible observational signatures. 
We find that stars with a major contribution from the turbulent pressure to the equation of state (i.e., \Pt/P>10\%) are confined to rather narrow band in the HR diagram (see Fig.\ref{fig:track}).
This high contribution is found in models where the HCZ is located very close to the surface, irrespective of their evolutionary stages.

When comparing the maximum fraction of \Pt within the stellar models to observationally derived macroturbulent velocities, we find a strong correlation between these two quantities, analogous to the finding of \citet{2015Grassitelli} for OB stars. Given that a similar strong correlation is found both in massive OB stars, where the high turbulent pressure appears in the iron convective zone, and in low- and intermediate-mass main- and post main-sequence stars, where the highest contribution from turbulent pressure lies within the HCZ, a physical connection of both phenomena appears likely. 

This connection can be interpreted as a consequence of the local deviation from the hydrostatic equilibrium due to turbulent pressure fluctuations at a percent level.  We argue that turbulent pressure fluctuations could lead to the excitation of high-order, non-radial pulsations, as observed in $\gamma$ Dor stars and even in our Sun, which manifest themselves at the surface as macroturbulent broadening of spectral lines \citep{2009Aerts}. 
If indeed turbulent pressure fluctuations excite high-order modes, a connection with $\gamma$ Dor may be present as well, which could involve finite amplitude pressure fluctuations in the HCZ.

Our results provide additional strong evidence for the connection between large-scale surface motions and the contribution of turbulent pressure, and calls for a revaluation, possibly inspired by 3D simulations, of the role of convection dynamics on the excitation and damping of both coherent and stochastically excited oscillation modes and waves. Rigorous studies should be performed to investigate the physical origin of the possible connection between sub-surface turbulent pressure fluctuations, macroturbulent broadening, and, e.g., the transition between $\delta$ Scuti (or the classical Cepheid instability strip), $\gamma$ Dor, and stochastically-excited pulsators. 
 This study appears particularly timely given the wealth of stringent observational constraints on photometric variability delivered, e.g., by CoRoT and {\it Kepler}.

\begin{acknowledgements} LG is part of the International Max Planck Research School (IMPRS), Max-Planck-Institut f{\"u}r Radioastronomie, and Universities of Bonn and Cologne. LF acknowledges financial support from the Alexander von Humboldt foundation. We thank Timothy Van Reeth for sharing the spectra of the Kepler stars and Konstanze Zwintz for useful discussions.
\end{acknowledgements}

\vspace{-0.5cm}

%

%

%------------------------------------------------------
%\newpage
\begin{appendix}

\section{Effects of \Pt\ on a 1.5\Msun stellar model and derived stellar parameters}

\begin{table}[h]
\caption[ ]{\vsini\ and \vmac values derived for the analysed stars.}
\label{tab:vsini-vmac}
\begin{center}
\begin{tabular}{l|cc}
\hline
\hline
Star & \vsini & \vmac \\
     & [\kms] & [\kms]\\
\hline
HD\,167858    & 6.2$\pm$1.0 & 9.8$\pm$1.5 \\
KIC\,8378079  & 1.5$\pm$2.6 & 12.5$\pm$1.1  \\
KIC\,9751996  & 3.9$\pm$3.4 & 12.9$\pm$1.5    \\
KIC\,11754232 & 2.4$\pm$2.5 & 11.8$\pm$1.1    \\
\hline
\end{tabular}
\end{center}
\end{table}

In this section we discuss the comparison between tracks with and without the inclusion of \Pt and $\rm e_{turb}$ in the calculation of a 1.5\Msun stellar models. We emphasize that, different from massive stars \citep{2015Grassitelli}, the convective velocities for intermediate-mass stars do not exceed the local sound speed and no limitation of $v_c$ is necessary.

During the main-sequence phase (\Teff\,$\gtrsim$\,5800\,K), the two tracks do not present appreciable differences; as the tracks approach the terminal-age main-sequence, the model with \Pt\ presents a displacement on the order of $\approx$10\,K toward lower effective temperatures. A similar behavior is also found for the post-main-sequence evolution, corresponding to a difference in radius on the order of $\approx$1\% or less. In line with the results of \citet{2015Grassitelli} for the massive stellar models (including a 7\Msun stellar track), we consider the effects of turbulence on the hydrostatic models to be small, and do not directly include them in the calculations of the stellar tracks in Fig.\ref{fig:track}.
\begin{figure}
\resizebox{\hsize}{!}{\includegraphics{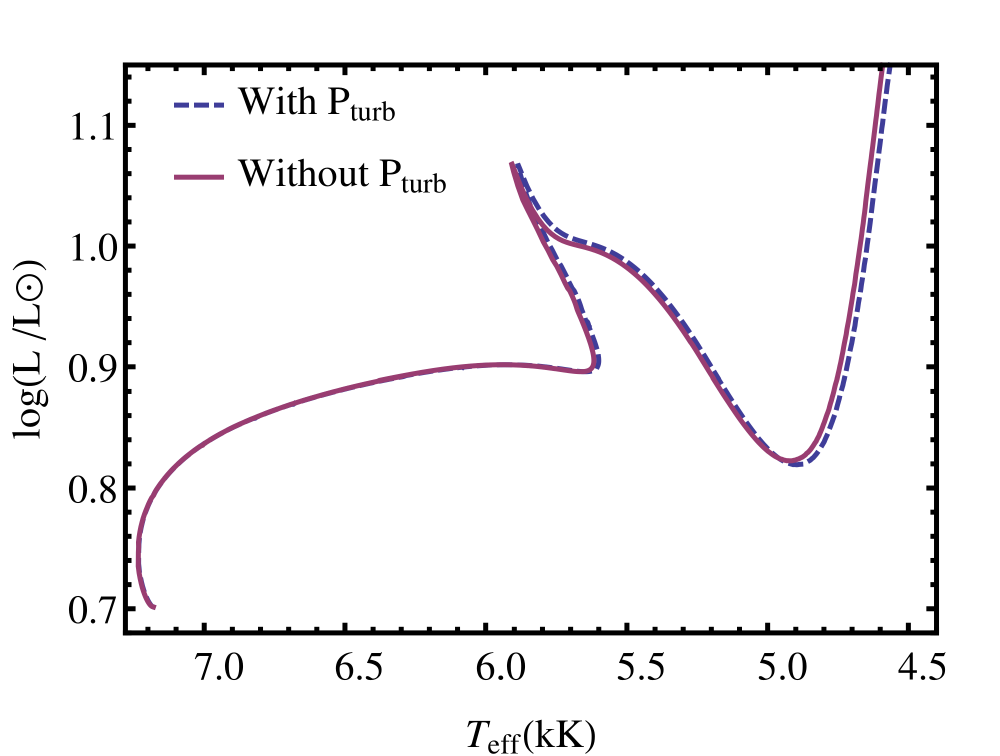}}
\caption{HR diagram showing evolutionary tracks of a 1.5\Msun model with (blue dashed) and without (purple continuous) turbulent pressure.}
\label{fig:1.5}
\end{figure}

\section{Convective timescales and mixing length}\label{app:conv_times}
\begin{figure}
\resizebox{\hsize}{!}{\includegraphics{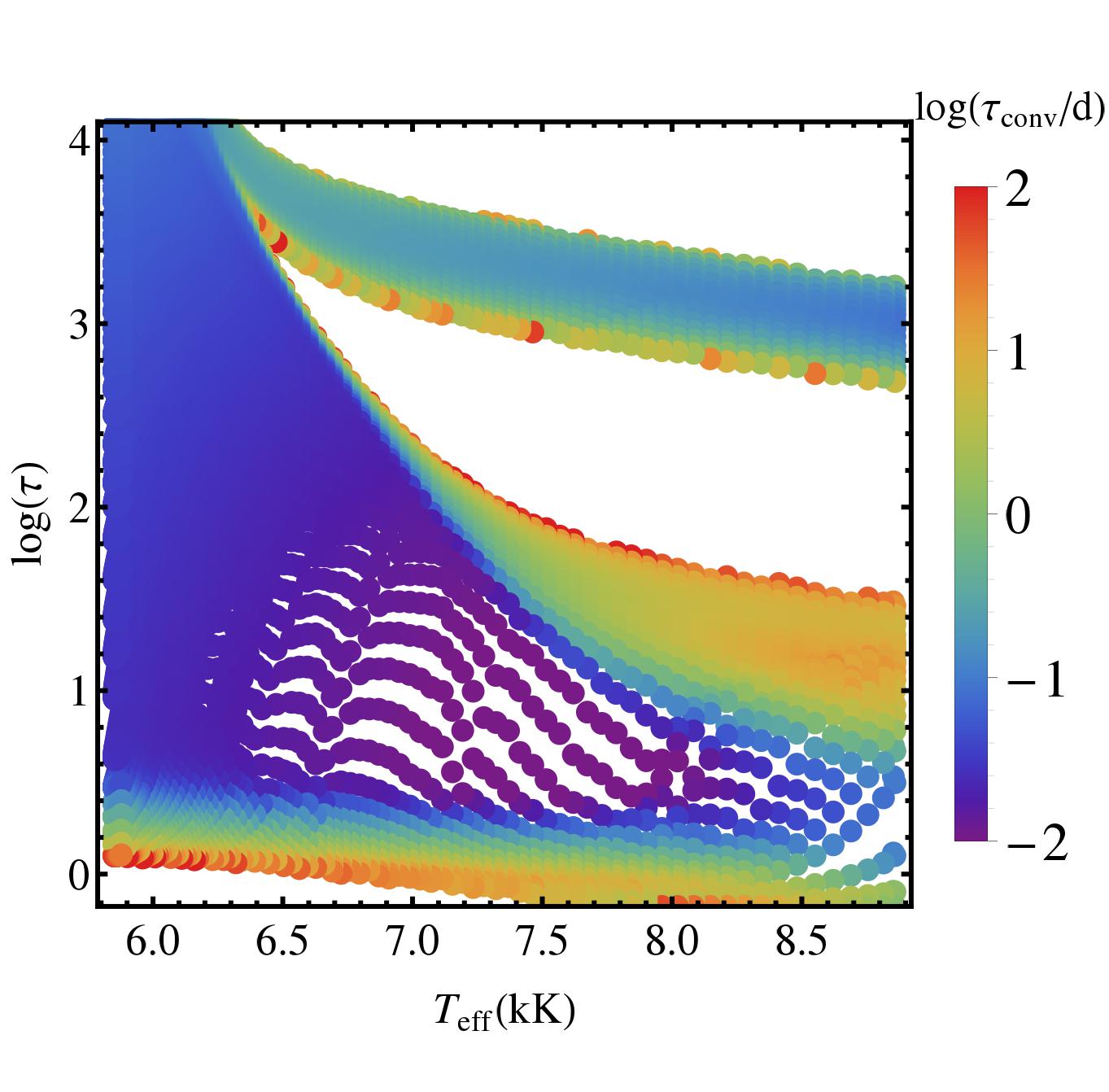}}
\caption{Convective timescale (color coded) as a function of the effective temperature and optical depth throughout part of the evolution of a $1.9\Msun$ model (same as Fig.\ref{fig:tefftau}). In the center of the hydrogen convective zone, where we find the higher contribution from \Pt, the convective timescales are $\approx 10^{-2}$ days, while at the edges of the hydrogen convective zone $\tau_{conv}$ increases up to $\approx 10^1-10^2$ days.}
\label{fig:tauconv}
\end{figure}

The results of an estimate of the convective timescales are shown in Fig.~\ref{fig:tauconv}. The convective timescale $\tau_{conv}$ has been computed as
\begin{equation}
\tau_{conv}=\frac{\alpha H_p}{v_c}
,\end{equation} 
where $H_p$ is the local pressure scale height. At the center of the hydrogen convective zone, where we obtain the highest convective velocities, the timescale for convection is on the order of five\,minutes. At the upper border instead (i.e., at the convective boundary close to the surface) the timescale increases from 1 up to $\approx 10$ days or more. Thus the convective timescales vary by $\approx 4$ orders of magnitude within the last mixing length of the convective zone. This makes a direct comparison with the observed periods difficult. Numerical simulations are necessary in order to show whether observed mode lifetimes are compatible with stochastic excitation. 

 The absolute values given here and in Sect.3 depend upon the adopted value of the mixing-length parameter $\alpha$. For $\alpha$\,=\,1, the turbulent pressure fraction decreases to a maximum of 5\% as a consequence of smaller convective velocities, while an increase of $\alpha$ (e.g., $\alpha$\,=\,1.8) leads to transonic convective velocities. Changing $\alpha$ also changes the position of the local peak of \Ptmax, e.g., the position of the peak decreases by 300\,K when adopting $\alpha$\,=\,1. 
%  
%-------------------------------------------
%-------------------------------------------
%
\end{appendix}
%------------------------------------------------------

\end{document}